\documentstyle[twocolumn,aps,epsf]{revtex}
\begin{document}
\newcommand{\beq}{\begin{equation}}
\newcommand{\eeq}{\end{equation}}

\title{SIMPLE MODELS OF PROTEINS WITH REPULSIVE NON-NATIVE CONTACTS}

\author{Mai Suan Li and Marek Cieplak} 

\address{ Institute of Physics, Polish Academy of Sciences,
Al. Lotnikow 32/46, 02-668 Warsaw, Poland }
\address{
\centering{
\medskip\em
{}~\\
\begin{minipage}{14cm}
The Go model is extended to the case when the non-native contact
energies may be either attractive or repulsive.
The folding temperature is found to increase with the energy of
non-native contacts.
The repulsive non-native contact energies may lead to folding
at $T=0$ for some two-dimensional sequences and to reduction in complexity
of disconnectivity graphs for local energy minima.
{}~\\
{}~\\
{\noindent PACS numbers:  71.28.+d, 71.27.+a}
\end{minipage}
}}

\maketitle


Functionally useful proteins are sequences of amino acids that fold rapidly
under appropriate conditions (temperature range, acidity of the water
solution etc) into their native states commonly assumed to be their ground
state configurations\cite{Dill}. The dynamics of folding is akin to 
motion in a rugged free energy landscape\cite{Bryngelson} and it crucially
depends on two factors: the interactions between the amino acids
and the target conformations.
In this paper, we focus on the role of non-native contact energies
in the folding process.
We study this issue in a simple model which is an extension of the
so called Go model proposed by Go and Abe\cite{Go}

Specifically, we consider the standard
two-dimensional lattice model of 16 monomers.
Its Hamiltonian is as follows

\begin{equation}
H \; \; = \; \; \sum_{i<j} \, \alpha_{ij} \Delta_{ij} \; ,
\end{equation}
where $\Delta_{ij}=1$ if monomers $i$ and $j$ are in contact and
$\Delta_{ij}=0$ otherwise (monomers $i$ and $j$ are considered
to be in contact if they are seperated by one lattice bond and
$|i-j| \neq 1$). The quantity $\alpha_{ij}=-1$ if monomers $i$ and $j$
are in contact in the native conformation and  $\alpha_{ij}=\alpha$
otherwise. The sequence is thus defined by the native conformation.
We allow $\alpha$ to be attractive 
($\alpha < 0$) or {\em repulsive}
($\alpha > 0$).
In what follows $\alpha$ is assumed to be larger
than -1.0 so that the native state is guaranteed to
be a maximally compact conformation.
In the original Go model\cite{Go} $\alpha=0$.
It should be noted that using model (1) one can monitor the effect of 
non-native contact energies by varying only
one parameter $\alpha$. Furthermore,
the two-dimensional model is simple enough to study the effect of the
target conformations on the folding dynamics.  

It should be noted that there have already been several studies
of models with repulsion in the non-native contacts:
a Gaussian model \cite{Shriva}, a designed model \cite{Gutin},
and the so called HP+ model
\cite{HP+}. The generalized Go model that we study here allows one to
vary the strength, and sign, 
of the non-native contacts relative to the native ones.

We focus on 4 target conformations shown in Fig. 1. There are 37 compact
conformations (for the two-dimensional 16-monomer chain one has 69 compact
$4\times 4$ conformations but for the Go-like model only 38
of them remain different due to the end-to-end reversal symmetry
and one is not accessible kinetically) 
which may act as the native conformations. Among
these $S_1$ and $S_4$  shown in Fig. 1 are the  two fastest folders at 
$T \neq 0$, whereas $S_2$ and $S_3$, also shown in Fig. 1, have intermediate
folding properties.

\begin{figure}
\epsfxsize=3.2in
\centerline{\epsffile{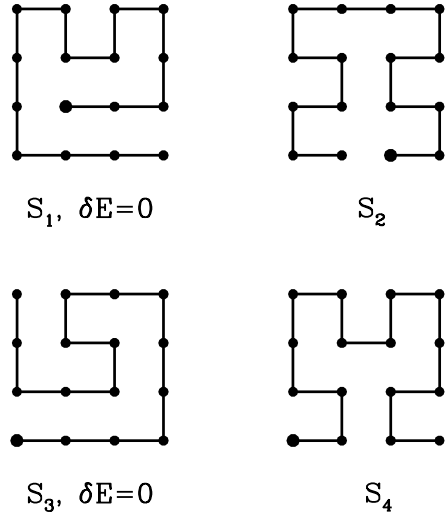}}
\caption{Four sequences studied in this paper
. $S_1$ and $S_3$ have $\delta E=0$.}
\end{figure}

It should be noted that the repulsive non-native contact interaction are
expected to improve the foldicity\cite{Shriva,Gutin,HP+} because they
restrict the 
size of the relevant phase space. In our case, however, this effect becomes
so dramatic that for $\alpha > 0.1$
sequences $S_1$ and $S_3$ can fold with a finite $t_{fold}$ even at $T=0$. 
Such an exotic phenomenon has been also observed in the HP model for some
13-monomer chains \cite{Chan1}  so it is not restricted to the Go-like
sequences that we study. Here, we study how this arises as a function
of $\alpha$ and show how do folding characteristics, i.e. characteristic 
temperatures and folding times, depend on $\alpha$.
The ability of $S_1$ and $S_3$ to fold at $T=0$
may be partly understood by the fact that for these sequences
the repulsive non-native interactions reduce the number of local minima
by two orders of magnitude compared to the case of $\alpha < 0$.
Furthermore, repulsive interactions dramatically affect partitioning
of the phase space into regions associated with the local energy minima.
We demonstrate this by
using the disconnectivity graph technique \cite{Karplus}
and show, in particular, that connectivities to the folding funnel
become simplified significantly.


\begin{figure}
\epsfxsize=3.2in
\centerline{\epsffile{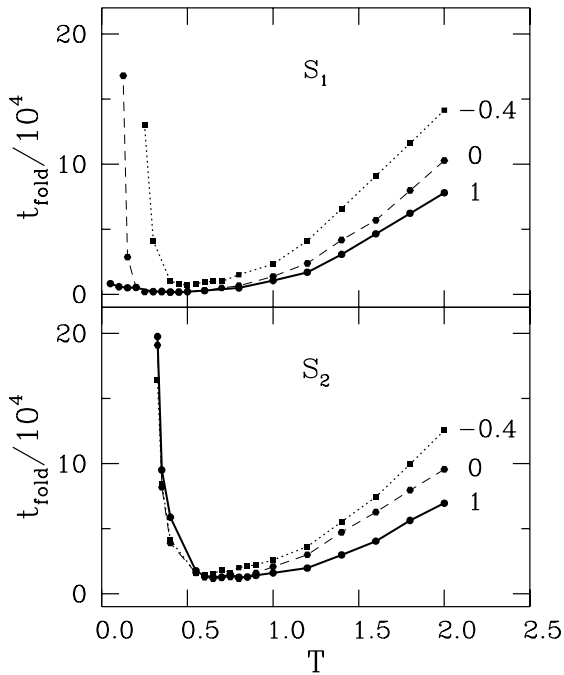}}
\caption{Temperature dependence of $t_{fold}$ for $S_1$
and $S_2$ and for 3 selected values of $\alpha$,
$\alpha = -0.4, 0$ and 1. The results are based on 2 - 4 batches,
each containing 200 trajectories from initial random conformations.}
\end{figure}

We have found that the folding temperature, $T_f$, increases with $\alpha$.
This result agrees with that of Camacho\cite{Camacho} for an
effectively zero-dimensional model\cite{Shakhnovich1}. 
We obtain it, however,
not only by the numerical calculations  for the
lattice model but also by the analytical argument.
The minimum folding time, $t_{min}$, defined at the temperature
where the folding is fastest has found to decrease with $\alpha$ and
it gets saturated for $\alpha \rightarrow \infty$.


The folding dynamics of a chain is studied by a Monte Carlo procedure
that satisfies the detailed balance condition\cite{Malte}, and was
motivated by the studies presented in Ref.\cite{Chan1,Chan}. The dynamics
allows for single and two-monomer (crankshaft) moves. For each
conformation of the chain one has $A$ possible moves and the maximum
value of $A$, $A_{max}$, is equal to $A_{max}=N+2$. In our 16-monomer case
$A_{max}=18$. For a conformation with $A$ possible moves, probability
to attempt any move is taken to be $A/A_{max}$ and probability not to do
any attempt is 1-$A/A_{max}$ \cite{Chan1,Chan}. In addition, probability
to do a single move is reduced by the factor of 0.2 and to do
the double move - by 0.8 \cite{Chan,Chan1}.
The attempts are rejected or accepted as in the standard Metropolis
method.
The folding time is equal to the total number of Monte Carlo attemps
divided by $A_{max}$.

We have carried out the Monte Carlo simulations to determine the dependence
of the median folding time,
$t_{fold}$, on $T$ and $\alpha$. The results for $S_1$ and $S_2$
are shown in Figure 2.
For each temperature, $t_{fold}$ is obtained based on 200 independent runs
starting from random configurations. The results are averaged over 2 - 4
batches, of 200 trajectories each.

\begin{figure}
\epsfxsize=3.2in
\centerline{\epsffile{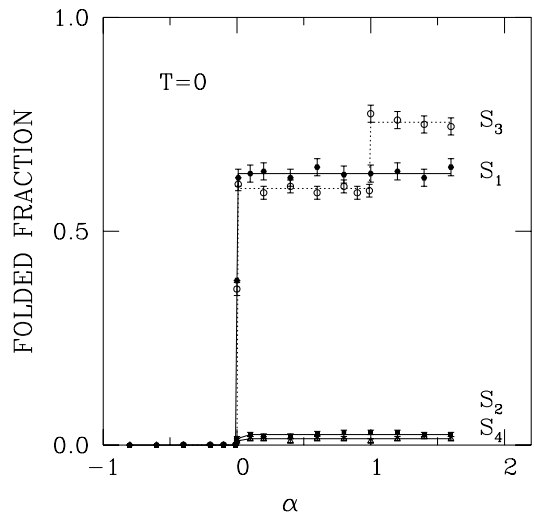}}
\caption{Dependence of the folded fraction on $\alpha$ at $T=0$ for
4 sequences shown in Figure 1. The results are averaged
over 4 - 6 simulations, each
corresponding to 200 trajectories.}
\end{figure}

For sequence $S_2$ the standard U-shape\cite{Socci} for the temperature 
dependence of $t_{fold}$ is observed for all values of $\alpha$. In 
other words, no qualitative change occurs if the non-native contact energies
change from attractive to repulsive.
In the case of sequence $S_1$, however, for  $\alpha=1$ 
the standard U-shape disappears suggesting that the glass
transition temperature $T_g$ which is operationally
defined  as the value of the temperature at which the median time
is  equal to some cut-off value (usually this cut-off value is chosen to
be 300000 Monte Carlo steps for the two-dimensional 16-monomer 
chain\cite{Socci}) becomes zero. In the standard scenario, 
at low temperatures the system may get trapped in some local minima
and the folding process becomes extremly
slow. The folding time is then governed by
the Arrhenius law, $t_{fold} \sim \exp(\delta E/T)$\cite{Shakhnovich,Malte},
where $\delta E$ is the energy barrier energy (at $T=0$ one has
$t_{fold} \rightarrow \infty$).
Thus, the absence of the $U$-shape dependence
suggests that the energy barrier $\delta E=0$.

\begin{figure}
\epsfxsize=3.2in
\centerline{\epsffile{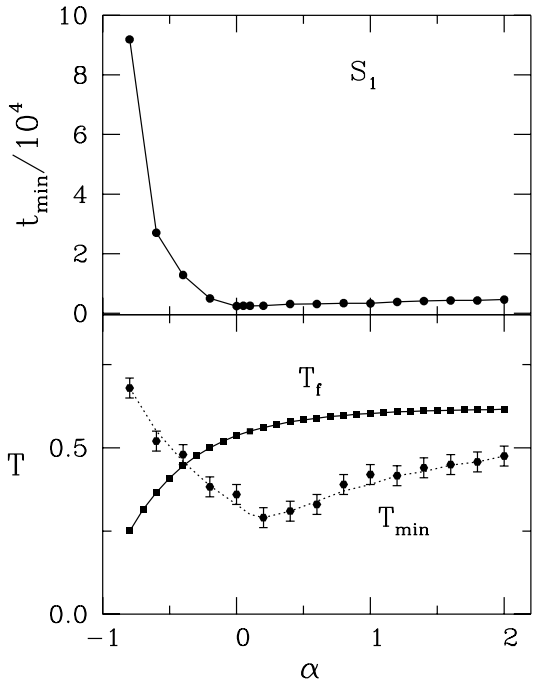}}
\caption{Dependence of $T_f$, $T_{min}$ and $t_{min}$ for
sequence $S_1$
on $\alpha$. The results are averaged over 4 batches, each containing 200
trajectories.}
\end{figure}

In order to know whether $\delta E$ is exactly zero or not one has to study
the folding at $T=0$.  At zero temperature the system gets trapped in some
local minimum or in the native state. If the fraction of the trajectories
from random conformations
fold into the native state is bigger than 50$\%$, then the chain
is said to be folded and  $\delta E=0$.

The fraction of folded trajectories is shown
in Fig. 3 for $S_1$, $S_2$, $S_3$ and $S_4$.
For $S_1$ and $S_3$ this fraction becomes bigger
than $50\%$ for $\alpha > \alpha _c$, where $\alpha _c \approx 0.1$.
Sequences $S_2$ and $S_4$ have $\delta E \neq 0$ for any value of $\alpha$.
It is interesting to mention that $S_4$ folds even faster than $S_1$ at
$T \neq 0$ but its foldicity becomes much worse at $T=0$.
Furthermore, the folding rate of $S_3$ at $T \neq 0$ is slower than for the
three other sequences and yet $S_3$  can fold at $T=0$. Thus
the geometry of the native targets has a dramatic effect
on the folding at $T=0$. Among the 37 maximally compact $4\times4$ structures
it is only $S_1$ and $S_3$ that do not obey the Arrhenius law.

\begin{figure}
\epsfxsize=3.2in
\centerline{\epsffile{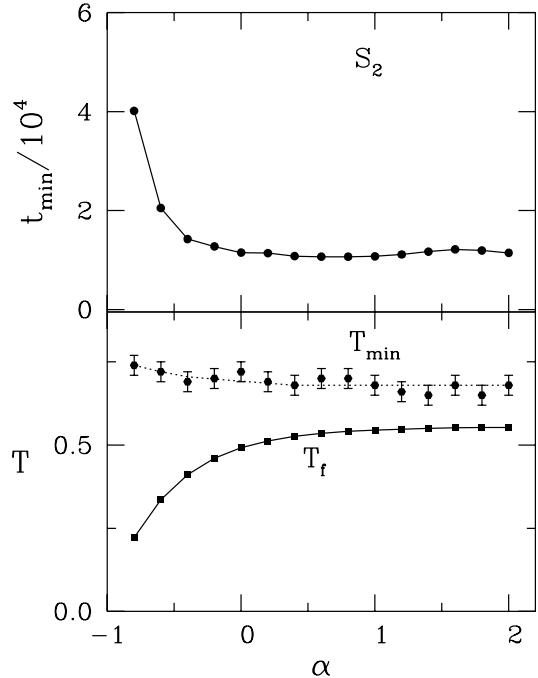}}
\caption{The same as in Figure 4 but for
sequence $S_2$.}
\end{figure}


We study the dependence of $T_f$ on $\alpha$ for
two typical sequences $S_1$ and $S_2$ shown in Fig. 1.
The total number of conformations of the 16-monomer chain is 
only 802075\cite{Lau,Dinner} and
is amenable to exact enumeration. This allows for an exact evaluation of the
equilibrium parameters such as the folding temperature $T_f$. The latter
is defined as a temperature at which the probability of occupancy of
the native state is $1/2$. The results for sequences $S_1$ and $S_2$
are shown in Fig. 4 and 5.
Obviously, $T_f$ increases with $\alpha$ but this dependence
gets weaker for larger values of $\alpha$.

The increase of $T_f$ with $\alpha$ may be understood in the following 
simple way. Let $P_{\Gamma _0}(\alpha)$ be the probability of occupying
the native state of conformation $\Gamma _0$. Then
\begin{equation}
P_{\Gamma _0}(\alpha) \; \; = \; \; 
\frac{\exp^{-\beta E_{\Gamma _0}}}{\sum_{\Gamma}
\exp^{-\beta\alpha n - \beta E'_{\Gamma}}} \; \; ,
\end{equation}
where $E_0$ is the energy of the native state, $n$ is the number of
non-native contacts and $E'_{\Gamma}$ is the part of energy in conformation
$\Gamma$ which corresponds to the native contacts.
Then
\begin{equation}
\frac{\partial P_{\Gamma _0}}{\partial \alpha} \; \; \geq
\; \; \beta <n>_T \; \geq 0 \; \; ,
\end{equation}
where $<n>_T$ is the average number of non-native bonds at temperature $T$.
So the probability of being in the native state cannot decrease  with $\alpha$. 
$T_f$, therefore, should increase with increasing $\alpha$ and 
then become $\alpha$-independent.

\begin{figure}
\epsfxsize=3.2in
\centerline{\epsffile{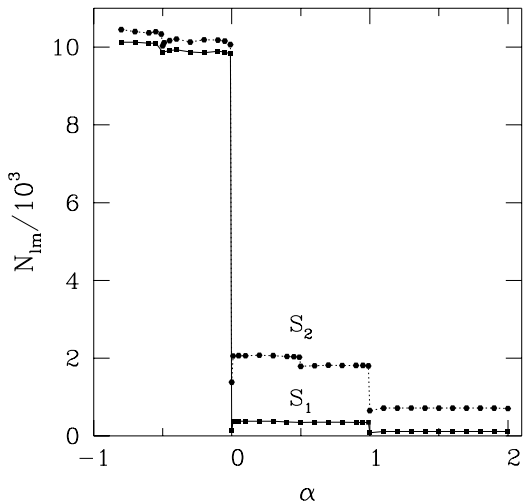}}
\caption{Dependence of $N_{lm}$ on $\alpha$ for
$S_1$ and $S_2$.}
\end{figure}

It should be noted that
in Camacho's model \cite{Camacho} $T_f$
was found to increase with
$\alpha$ ($\alpha<0$) linearly. Our results presented in Figures 4 and
5 show that the region of $\alpha$ where one can observe the linear
dependence is rather narrow. Such region becomes much wider, for example,
in the case of the 27-monomer chain in three dimensions (the results are
not shown). Overall, the results for $T_f$, shown in Fig. 4 and 5,
demonstrate that the non-native contact repulsive energies improve
both the thermodynamic stability 
and dynamical characteristics of folding.

We now focus on the $\alpha$ dependence of 
the temperature at which the folding time is minimal, $T_{min}$.
The results for $S_1$ and $S_2$ are shown
in Fig. 4 and Fig. 5. 
For positive values of $\alpha$, $T_f$ and $T_{min}$ are 
comparable for both 
sequences and they should be good folders\cite{Socci}.
For large negative values of $\alpha$, $T_{min}$ is bigger than $T_{f}$
but $S_1$ and $S_2$ remain good folders by the 
Thirumalai-Camacho criterion\cite{Thirumalai,Camacho1} -
the peaks of the structural susceptibility and
the specific heat coincide for this interval of $\alpha$.
(For $\alpha=0$, the coincidence of the peaks 
is, in fact, a general feature of the Go models because the proximity
to the native state and occurence of rapid changes in an average energy as a 
function of $T$ are both controlled by establishment of the same
native contacts).

\begin{figure}
\epsfxsize=3.2in
\centerline{\epsffile{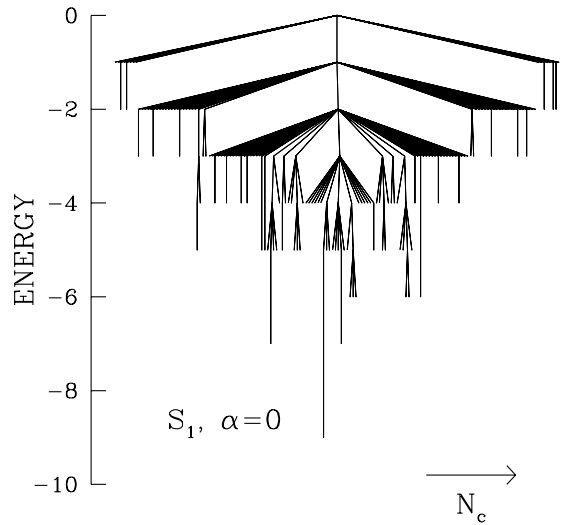}}
\caption{The disconnectivity graph for $S_1$ and $\alpha =0$. 
$N_c$ is a symbolic notation for the index labelling the local energy minima.}
\end{figure}

Fig. 4 and 5  also show the dependence of the minimal folding time,
$t_{min}$, on $\alpha$.
The dependence seems to saturate at large values of $\alpha$.
If one extends the Camacho result to positive $\alpha$
then $t_{min}$ should decrease with $\alpha$ exponentially\cite{Camacho}.
Our results suggest that such conclusion is valid only for $\alpha < 0$
but not for $\alpha>0$. 
Thus, the repulsive non-native contact energies make
the polypeptide chain to fold faster which is similar
to what has been observed in Ref.\cite{Shriva,HP+}.

In order to understand why the repulsion improves the folding so much
we study the dependence of number of local minima, $N_{lm}$, on $\alpha$.
The results for $S_1$ and $S_2$ are shown in Fig.6.
$N_{lm}$ of $S_1$ is found to be smaller than for $S_2$.
Clearly, the number
of local minima strongly depends on $\alpha$ and  for 
$\alpha >0$ it become by 2 orders
of magnitude smaller than that for $\alpha <0$.
So, the positive values of $\alpha$ make the energy landscape less rugged
and the folding dynamics get faster.

\begin{figure}
\epsfxsize=3.2in
\centerline{\epsffile{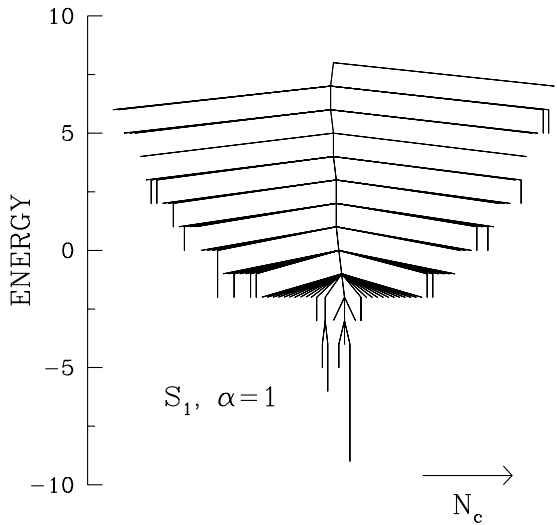}}
\caption{The same as in Fig. 7 but for $\alpha =1$.
}
\end{figure}

In order to get more insight into the nature of the energy landscapes
in the models studied here, we use the
disconnectivity graph technique which maps the potential energy
surface onto the set of local minima \cite{Karplus,Wales}.
The technique involves checking what local energy minima are
connected by pathways (sets of moves that are allowed kinetically)
that do not exceed a given total threshold energy.
For each value of this energy the minima
are divided into disconnected sets of mutually accessible minima
separated by barriers.
The local minima which share 
the lowest
energy threshold are joined at a node by lines and are called a basin
corresponding to the threshold.
The procedure of construction is stoped when one gets only
one basin for all of the minima.
For $S_1$ the number of local minima is equal to 152 and 81
for $\alpha =0$ and $\alpha =1$, respectively. The construction
of the graphs, therefore,
may be done exactly.
The disconnectivity
graphs obtained for $S_1$ and $\alpha =0$ and $\alpha =1$
are shown in Fig. 7 and Fig. 8, respectively. For both values
of $\alpha$ sequence $S_1$ is a good folder and consequently,
the structure corresponding to a folding funnel is clearly visible. 
However, the funnel for the repulsive case
of $\alpha =1$ has a significantly less complex pattern than for
$\alpha =0$. Moreover, the repulsive non-native interactions  
reduce the number of the local energy minima that have 
links to the native state below any predetermined energy threshold.
Thus, the comparison of the
disconnectivity graphs also shows that the repulsion may
facilitate the folding substantially.

\begin{figure}
\epsfxsize=3.2in
\centerline{\epsffile{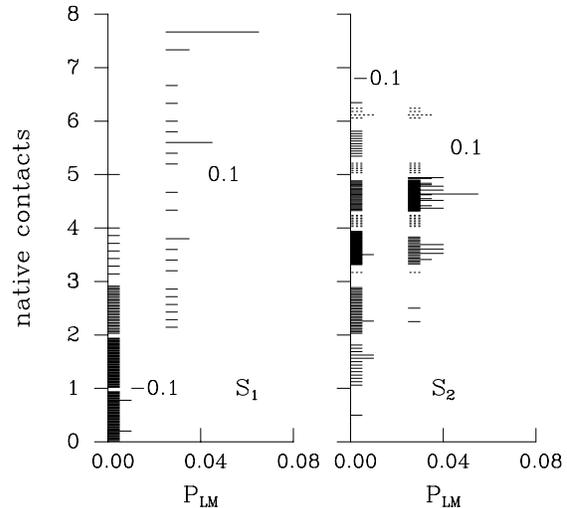}}
\caption{The density of the ($V$-shape) local minima,
$P_{LM}$, in which the system
gets trapped at $T=0$ is plotted versus the number of native contacts
for $S_1$ and $S_2$.
We choose $\alpha = 0.1$ and $\alpha = -0.1$. The results are obtained
for the batch of 200 trajectories. In the case of $S_1$ the local minima are
different for $\alpha =-0.1$ and $\alpha =0.1$.
For $S_2$ there are 19 common local minima which are marked by dotted lines.}
\end{figure}

We now address the question of what happens with trapped local minima 
when $\alpha$ changes from negative to positive values. Fig. 9 shows the
histogram of the local minima  for $\alpha=-0.1$
and 0.1 for sequence $S_1$ and $S_2$. Interestingly, for $S_1$
none of the local 
minima obtained
for $\alpha=-0.1$ appears for $\alpha=0.1$. The situation
changes dramatically for $S_2$ for which 19 
local minima are common for both  
$\alpha=-0.1$ and $\alpha=0.1$. 
Our results suggest that for the
sequence with $\delta E =0$ the local
minima in which the chain is trapped at the negative values of $\alpha$
are effectively avoided if $\alpha$ is changed to positive values.



In conclusion, we state that for
the simple extended  Go model
$T_f$ ($t_{min}$) increases (decreases) with the non-native contact 
energy and it gets saturated for large values of $\alpha$.
The complexity of the disconnectivity trees becomes reduced on making
$\alpha$ more and more repulsive so that  some two dimensional
sequences may even lose the Arrhenius like behavior of $t_{fold}$
at low temperatures.
It would be interesting to determine whether there are
any three-dimensional Go-like sequences that fold even at $T=0$.
Another important question is
what kind of an effective parameter $\alpha$,
or its conceptual equivalent, characterizes real proteins. 

We thank Jayanth R. Banavar, P. Garstecki
and T. X. Hoang for many useful discussions.
This work was supported by KBN (grant No. 2P03B-025-13).

\end{document}